# Dual-Wavelength Lasing in
# Quantum-Dot Plasmonic Lattice Lasers

*Jan M. Winkler, Max J. Ruckriegel, Henar Rojo, Robert C. Keitel, Eva De Leo, Freddy T. Rabouw,[†] and David J. Norris[*]*

Optical Materials Engineering Laboratory,

Department of Mechanical and Process Engineering,

ETH Zurich, 8092 Zurich, Switzerland

ABSTRACT. Arrays of metallic particles patterned on a substrate have emerged as a promising design for on-chip plasmonic lasers. In past examples of such devices, the periodic particles provided feedback at a single resonance wavelength, and organic dye molecules were used as the gain material. Here, we introduce a flexible template-based fabrication method that allows a broader design space for Ag particle-array lasers. Instead of dye molecules, we integrate colloidal quantum dots (QDs), which offer better photostability and wavelength tunability. Our fabrication approach also allows us to easily adjust the refractive index of the substrate and the QD-film thickness. Exploiting these capabilities, we demonstrate not only single-wavelength lasing but dual-wavelength lasing via two distinct strategies. First, by using particle arrays with rectangular lattice symmetries, we obtain feedback from two orthogonal directions. The two output wavelengths from this laser can be selected individually using a linear polarizer. Second, by adjusting the QD-film thickness, we use higher-order transverse waveguide modes in the QD film to obtain dual-wavelength lasing at normal and off-normal angles from a symmetric square array. We thus show that our approach offers various design possibilities to tune the laser output.

Nanometer-sized metallic particles support plasmonic resonances at visible wavelengths.[1] When such particles are arranged in a two-dimensional lattice, the localized surface plasmon resonances (LSPRs) from each particle hybridize with in-plane-propagating optical modes. The resulting surface lattice resonances (SLRs)[2–4] combine strong electric-field confinement near the plasmonic particles with high quality factors of the photonic modes. These particular properties turn SLRs into a useful platform for studying light–matter interactions. Emitters weakly coupled to plasmonic lattices show enhanced radiative decay rates and modified emission directionality.[5–8] When the interactions reach the strong-coupling regime,[9–13] condensation[14] and exciton-polariton lasing[15] have been observed.

Furthermore, arrays of metallic particles have emerged as promising resonators to provide feedback in on-chip nanolasers.[16–19] Most recent realizations of plasmonic-array lasers used dye molecules dissolved in a host liquid as the gain material.[20–22] The use of liquid gain media was motivated by two reasons. First, Brownian diffusion in the liquid has the effect of replenishing bleached dye molecules in regions of intense optical pumping, thus minimizing the effects of decomposition of the gain material. Second, by tuning the refractive index of the solvent, one can conveniently index-match the sub- and superstrates of the plasmonic array, which is required to excite SLRs.[4] However, integrating a liquid gain material into a device envisioned for on-chip applications is challenging. As an alternative, a dye-doped polymer layer with a refractive index intentionally exceeding that of the substrate has been exploited.[23,24] There, feedback was provided through lattice modes that arise due to coupling of the plasmonic particles to the waveguide modes supported by the high-index polymer layer. Nevertheless, photostability remains a challenge for any system using organic gain materials.

Colloidal semiconductor nanocrystals offer a photostable and wavelength-tunable alternative to organic dyes.[25–27] Dispersed in volatile solvents, these particles, also known as quantum dots (QDs), retain the advantageous properties of molecules such as solution-processability and scalability. Assembled into dense films through simple drop-casting or spin-coating, QDs have served as the gain material in distributed-feedback lasers[28] and ring lasers.[29] However, although control over the spontaneous emission of QDs with plasmonic-particle arrays has been demonstrated,[30,31] these materials have not yet been explored as the gain material in plasmonic-lattice lasers.

Here, we integrate CdSe/CdS/ZnS core/shell/shell QDs with arrays of plasmonic Ag disks for single- and dual-wavelength lasing at visible wavelengths. We adapt the method of template stripping[32] to create plasmonic lattices on epoxy substrates with variable refractive index. QD films of controllable thickness are drop-cast as the gain medium onto these arrays. We can tune the feedback conditions of our plasmonic lattices not only by changing their periodicity but also by varying the layer stack surrounding the Ag disks, i.e. the QD-film thickness and/or the optical constants of the epoxy. With a thin QD film on top of a square Ag-disk lattice, we obtain a single-wavelength laser that emits normal to the lattice plane. We can expand the operation of our QD lasers from single- to dual-wavelength via two separate approaches. First, we distort the square into a rectangular array and thus obtain feedback at two tunable frequencies, each determined by one of the lattice periodicities. Because the resulting lasing wavelengths possess perpendicular polarization we can control their far-field emission using a linear polarizer. In a second approach, we return to the square-lattice geometry but now increase the QD-film thickness and the substrate refractive index. For sufficiently thick QD films, the stack supports both zero- and higher-order waveguide modes, which are coupled through diffraction by the plasmonic lattice. In addition to the fundamental normal-angle lasing, feedback in these coupled modes can result in off-normal laser emission at a second wavelength. Our findings highlight the potential of QDs as a gain material for plasmonic-lattice lasers and provide guidance for engineering the feedback conditions of such devices.

We begin by discussing the plasmonic-lattice resonator without QDs (Figure 1). We fabricated Ag-disk lattices on transparent substrates using a template-stripping-based approach adapted from previous reports (Figure 1a).[33,34] First, we prepared a reusable $SiO_2$/Si template patterned with an array of shallow circular depressions by combining electron-beam lithography and selective etching of the 50-nm-thick $SiO_2$ layer. To reduce adhesion during the subsequent template-stripping step, we covered the $SiO_2$ layer with a self-assembled monolayer of octadecyltrimethoxysilane (step i, Figure 1a; see also Section S1 in the Supporting Information).[35] Then, 25 nm of Ag were thermally evaporated onto the template at a high deposition rate (step ii, Figure 1a).[36] Adhesive tape was used to remove the Ag deposited onto the SiO$_2$ surface while the metal inside the circular depressions remained (step iii, Figure 1a and Figure S1 in the Supporting Information). Finally, the Ag disks were stripped out of the etch pits using an epoxy cured with ultraviolet (UV) light (step iv and v, Figure 1a). The epoxy was also carefully chosen to have a suitable refractive index. In contrast to previous reports where the template could only be used once,[37] we can reuse our SiO$_2$/Si template after appropriate cleaning (see Section S1 in the Supporting Information). Hence, nanofabrication is only required once to make the initial template, from which many copies of the same particle array (but made from potentially different metals[38] and substrate refractive indices) can be produced.

A scanning electron micrograph (SEM) in Figure 1b shows a portion of a resulting template-stripped 80×80 µm$^2$ plasmonic Ag square array, with a particle diameter of 60 nm and a lattice spacing (pitch) of 370 nm. Other lattice geometries and disk diameters can be realized by adapting the template within the resolution limits of electron beam lithography. However, to ensure that the lattices can be stripped reliably, the disk aspect ratios (height to lateral dimension) should be smaller than one.

A template-stripped array of Ag disks supports SLRs that depend on the geometry and the dielectric environment of the lattice.[3,4] Any two-dimensional periodic structure scatters incoming light with momentum $\vec{k}_{\text{inc}}$ into diffraction orders $[i, j]$ with momentum $\vec{k}_{i,j}$ by altering the in-plane momentum according to

$$\vec{k}_{\text{inc},\parallel} + i\vec{G}_1 + j\vec{G}_2 = \vec{k}_{i,j,\parallel}, \text{ with } i,j = 0, \pm 1, \pm 2, \ldots \qquad (1)$$

where $\vec{G}_1, \vec{G}_2$ are the reciprocal lattice vectors of the array and $\vec{k}_{\text{inc},\parallel}$, $\vec{k}_{i,j,\parallel}$ are the in-plane wave-vector components of the incoming light and the $[i, j]$ diffraction order, respectively. Special diffraction events occur when the diffracted wave travels in plane. This arises for incoming photons with energy $\hbar\omega$ when the so-called Rayleigh condition[39] is fulfilled:

$$|\vec{k}_{i,j,\parallel}| = |\vec{k}_{i,j}| = n\frac{\omega}{c} \qquad (2)$$

where $n$ is the refractive index of the dielectric medium surrounding the lattice. Combining eq 1 and eq 2, we obtain the Rayleigh condition for a particular lattice as a function of the in-plane momentum $\vec{k}_{\text{inc},\parallel}$ of the incoming light. We refer to this as the empty-lattice dispersion, since the effect of the Ag disks on the refractive index $n$ in eq 1 is neglected. For light propagating in the $xz$-plane (see schematic in Figure 1c) that is incident onto a square array with a pitch $a_{x,y} = 385$ nm, this empty-lattice dispersion relation is plotted as white dashed lines in Figure 1d. The left and the right section of the Figure correspond to an array surrounded by $n = 1.56$ and $n = 1.7$, respectively. The two linear features labeled with $[i, j] = [1,0], [-1,0]$ represent the condition where light incident with $\vec{k}_{\text{inc},\parallel} = (k_x, 0)$ is diffracted by $\pm \vec{G}_1 = \pm \vec{G}_x = \pm (2\pi/a_x, 0)$ into an in-plane propagating mode. The degenerate parabolic feature $[i, j] = [0, \pm 1]$ is due to scattering of the incoming wavevector by $\pm \vec{G}_2 = \pm \vec{G}_y = \pm (0, 2\pi/a_y)$.

In our array, nanometer-sized Ag disks supporting LSPRs at visible wavelengths occupy every lattice site. Hence, in-plane-propagating optical modes excited on the empty-lattice dispersion interact with these plasmonic resonances and form SLRs.[3,4] A Fourier-imaging setup can be used to experimentally measure the dispersion relation of the hybrid SLRs (see Section S2 and Figure S2 in the Supporting Information for details). The particle array is excited with white light and the Fourier image of the transmitted signal is then projected onto the entrance slit of an imaging spectrograph. By aligning the slit along the $x$-axis of our sample and dispersing the signal, we obtain transmission spectra resolved along $k_x$ with $\vec{k}_{\text{inc},\parallel} = (k_x, 0)$.

The color map on the left side of Figure 1d shows the experimentally obtained transmittance of a plasmonic square lattice ($a_{x,y} = 385$ nm) that consists of disks with diameter $d = 70$ nm. The disks are completely embedded in epoxy of $n = 1.56$. This is achieved by the fabrication procedure described above, followed by overcoating with epoxy (step vi, Figure 1a). In the transmission measurement we observe a broad, non-dispersive dip in transmittance at ~2.35 eV that corresponds to the LSPR resonance supported by the individual Ag disks. Moreover, the SLR band structure is revealed by points of low transmittance, indicating successful coupling of photons to SLR modes at the corresponding momentum and energy values. The band structure follows the empty-lattice dispersion (white dashed lines) but is shifted to lower energy due to photon–LSPR hybridization.[3,4] The linewidth of the SLR band at $k_x = 0$ is 64 meV [full-width-at-half-maximum (fwhm)] yielding a quality factor of ~35. Plasmonic lattices of similar geometry but fabricated with conventional electron-beam lithography showed comparable $Q$-factors.[4]

In contrast to the empty-lattice dispersion, the SLR band structure forms a stop gap at $k_x = 0$ and $E \approx 2.1$ eV. At the energy of the stop gap, counterpropagating SLRs with in-plane wavevectors $\vec{k}_{SLR} = \pm \vec{G}_x = (\pm 2\pi/a_x, 0)$ are backscattered into each other by coupling to the lattice periodicity with $[i,j] = [\pm 2,0]$ (see eq 1), while SLRs with $\vec{k}_{SLR} = \pm \vec{G}_y$ backscatter into each other via $[i,j] = [0, \pm 2]$. Standing waves then form due to constructive interference of these counterpropagating SLR modes, with their energies determining the upper and lower edges of the stop gap. As in distributed feedback lasers,[40] these standing waves can also give rise to lasing.[21,24] When this occurs, laser light will scatter from the standing-wave modes into free-space photons that propagate normal to the lattice plane ($\vec{k}_{inc,\parallel} = 0$) through diffraction by $1\vec{G}_x$ or $1\vec{G}_y$ (see eq 1). The energy at the center of the stop gap, and thus the resonance condition for lasing, can be estimated by combining eq 2 and the above-mentioned condition for backscattering of SLRs (referred to as the second-order Bragg condition):

$$E_{x,y} \approx \frac{hc}{n a_{x,y}} \tag{3}$$

where $h$ is Planck's constant.

According to eq 3, the energy where a lattice provides feedback is determined by the lattice spacing and the refractive index of the medium surrounding the array.[3,4] The modification of these parameters typically involves electron-beam lithography, to fabricate either a new array geometry or a plasmonic lattice on a substrate with a different refractive index. In contrast, our method of template stripping allows for convenient variation and tuning of the substrate refractive index without further lithography. A commercial library of UV-curable adhesives with refractive indices between 1.315 and 1.7 can be used. To demonstrate that the band structure and the associated resonance condition for lasing can be tuned via the refractive index, we evaporated Ag onto the same template ($a_{x,y} = 385$ nm, $d = 70$ nm) but embedded the Ag disks in an epoxy of $n = 1.7$. The color map on the right side in Figure 1d shows the measured SLR band structure for the resulting plasmonic lattice. Due to the higher refractive index, the LSPR appears at 2.2 eV and the SLR band structure is shifted by 196 meV (at $k_x = 0$) towards lower energy. The stop gap is now at 1.9 eV, which is in good agreement with eq 3 that predicts 1.90 eV.

The different SLR bands have distinct polarization properties that can be probed in the transmission measurement.[41] This will be relevant later for color selection in our dual-wavelength lasers. Figure 1e shows momentum-resolved transmission spectra of an array with the same geometry as shown in Figure 1d but illuminated with linearly (s or p) polarized light. S-polarized light with $\vec{k}_{\mathrm{inc},\parallel} = (k_x, 0)$ (see red polarization state in Figure 1c) selectively excites the $[i,j] = [1,0], [-1,0]$ SLR modes (see Figure 1e, left side). P-polarized light impinging from the same direction couples to the parabolic $[i,j] = [0, \pm 1]$ SLR features (see Figure 1e, right side).[41] The two sets of SLRs are orthogonally polarized because the "flat" Ag disks are mainly polarizable in-plane (*xy*-plane in Figure 1c). Because of their marginal height of 25 nm, their LSPR in the out-of-plane direction lies at significantly higher energy (not visible in the transmission maps of Figure 1d,e), and their out-of-plane polarizability—especially at the energy range of interest—is negligible. Hence, in-plane propagating photon modes can hybridize with the Ag disks only if they have transverse-electric polarization (i.e. electric field in plane).

Now that we understand the band structure of periodically arranged plasmonic particles in a homogeneous medium, we next integrate the QDs. We used CdSe/CdS/ZnS core/shell/shell nanocrystals synthesized according to a previously published recipe.[42] In liquid dispersion, these QDs exhibit a narrow Gaussian emission feature centered at 1.98 eV (625 nm) with a linewidth (fwhm) of 94 meV (30 nm). To fabricate the structure envisioned for lasing (see schematic in Figure 2a), we drop-cast QDs from a 9:1 (v/v) mixture of hexane:octane onto the plasmonic lattice. After complete evaporation of the solvents, a dense and continuous nanocrystal layer was formed. The gain material is thus not dispersed in a polymer matrix or a liquid. The thickness of the emitter film was controlled by changing the QD concentration in the dispersion, while maintaining the drop-cast volume. Finally, the device was sealed with a layer of epoxy that matches the refractive index of the epoxy substrate ($n_{\mathrm{epoxy}} = 1.56$).

In this structure, the plasmonic Ag disks are embedded inside a high-refractive-index QD film ($n_{QD} \approx 1.85$ at 630 nm)[43] that acts as a waveguide. In contrast to a lattice surrounded by a homogeneous refractive index where LSPRs hybridize with unbound photons that propagate in plane, the plasmon resonance of the Ag disks now couples to waveguide modes supported by the QD layer.[23] To estimate the band structure of such a device and its corresponding feedback condition for lasing, we have to consider the energy–momentum dispersion relation of the waveguided modes with an effective mode index $n_{eff}(\omega)$. This determines the conditions for coupling between free-space photons and in-plane propagating radiation as well as the coupling between counterpropagating waves. Therefore, we adapt the empty-lattice dispersion above (eqs 1 and 2) by replacing the refractive index $n$ in eq 2 by the frequency-dependent effective index $n_{eff}(\omega)$ of the waveguide mode of interest. Hence, the band structure of the plasmonic lattice is now governed by $n_{eff}(\omega)$, which in turn is determined by $n_{QD}$, $n_{epoxy}$, and the thickness of the QD layer.

Our approach to fabricate plasmonic lattices offers simple ways of controlling the effective mode index $n_{eff}(\omega)$ of the waveguide modes in the QD film. Figure 2a maps the range of effective mode indices that can be achieved with QD films ($n_{QD} = 1.85$) up to 500 nm thick surrounded by two semi-infinite epoxy layers ($n_{epoxy} = 1.56$).[44] For thin emitter films, only the fundamental $TE_0$ and $TM_0$ waveguide modes are supported, and their $n_{eff}$ is close to the refractive index of the epoxy layers. As the film is made thicker, $n_{eff}$ increases and approaches the refractive index of the QD film. For films exceeding the cutoff thickness of ~300 nm, higher-order $TE_1$ and $TM_1$ modes appear in addition to the fundamental waveguide modes. The $n_{eff}$ values for $TE_1$ and $TM_1$ also start off at $n_{epoxy}$ and evolve towards $n_{QD}$ for increasing film thickness. Thus, changing the thickness of the QD layer offers a simple method to adjust $n_{eff}$ between 1.56–1.85 and to select the number of modes supported by the waveguiding layer.

To fabricate a device for lasing, we aim for a 180-nm-thick QD film and deposit it onto a Ag-particle square lattice ($a_{x,y} = 390$ nm; $d = 70$ nm). This particular layer thickness ensures that only the fundamental $TE_0$ and $TM_0$ modes are supported by the waveguiding layer (orange region in Figure 2a). Because the QD film fluoresces, we can measure the band structure of the plasmonic lattice by exciting the emitters with a 385 nm light-emitting diode (LED) and detecting their momentum-resolved photoluminescence signal (Figure 2b). Upon photoexcitation, the QDs emit either into free-space or into one of the two waveguided ($TE_0$ or $TM_0$) modes. The free-space emission manifests itself as an undispersed broad emission band centered at 1.96 eV and reflects the Gaussian emission spectrum of the QDs. Due to energy transfer between QDs inside the film, the emission energy is red shifted by 20 meV compared to the signal obtained from QDs in liquid dispersion (see red and blue curves in the inset of Figure 2b).[45] Qualitatively similar to the SLR dispersion for the Ag square lattice in a homogenous medium, we observe in Figure 2b a band structure consisting of two linear $[i,j] = [1,0], [-1,0]$ bands that form a stop gap at ~1.91 eV and two parabolic $[i,j] = [0, \pm 1]$ bands centered around 1.92 eV. These bands result from QD emission into the $TE_0$ guided mode and subsequent diffraction by $1\vec{G}_{x,y}$ of the square lattice into free-space photons that our camera can record. The empty-lattice dispersion relation for $n_{\text{eff,TE}_0} = 1.665$ is plotted as white dashed lines at negative $k_x$ for comparison. A second faint set of bands is shifted towards higher energy and results from coupling of the QDs to the $TM_0$ guided mode with a higher effective mode index. These bands are best visible at $k_x = -0.6 \ \mu m^{-1}$, $E = 1.9$ eV and show a stop gap at ~1.95 eV. To better visualize the band structure, we plot the empty-lattice dispersion relation for $n_{\text{eff,TM}_0} = 1.625$ as white dotted lines for negative $k_x$. The polarization mismatch between the in-plane polarizable Ag disks and the mainly out-of-plane polarized $TM_0$ mode is responsible for the weak emission signal associated to the $TM_0$ bands. Based on the lattice spacing and the energy of the stop gap we can estimate the effective index of the $TE_0$ waveguide mode to be $n_{\text{eff,TE}_0} \approx 1.67$ at the emission wavelength of the QDs (see eq 3). Comparing this value with the data in Figure 2a we can extract a QD-film thickness of ~180 nm.

When the same structure is excited with ~340 fs laser pulses at 405 nm (65 μm spot size, 10 kHz repetition rate), lasing emission is observed. Figure 2c shows momentum-resolved emission maps as a function of pump fluence. Upon sufficient pumping a narrow lasing peak appears at 1.93 eV and the spectral linewidth narrows from 29 nm (linewidth of the QD-film emission) down to 1 nm (limited by the resolution of our spectrograph). The band structure evident in the middle panel of Figure 2c indicates that the lasing emission occurs at the high-energy edge of the stop gap. There, the $TE_0$ mode receives feedback through second-order Bragg diffraction and possesses lower loss than on the low-energy band edge.[24] At pump fluences right above threshold ($P_{th}$) we observe a lasing peak that is split in energy by ~2 meV. We discuss the potential origin of this peak splitting in Section S3 of the Supporting information and provide an emission map with higher energy resolution in Figure S3. In Figure 2d the laser output intensity is plotted as a function of pump fluence. A sudden increase in emission intensity is observed at a threshold of 0.16 mJ/cm$^2$, which is characteristic for lasing. The threshold fluence depends on the spectral overlap between the feedback conditions in the plasmonic lattice and the gain envelope of the QDs. We can improve this overlap using a geometry with $a_{x,y} =$ 380 nm; $d = 70$ nm, resulting in a stop gap at ~1.96 eV and a slightly lower threshold value of 0.11 mJ/cm$^2$ (see Section S3 and Figure S4 in the Supporting Information). However, it was not the goal of this work to optimize the device towards the lowest possible threshold values.

By measuring the Fourier image of the lasing signal without spectrally dispersing it, we retrieve the full 2D in-plane momentum map. Above threshold the momentum-resolved laser emission of our sample shows a cross with a single bright spot in the center at $k_x = k_y = 0$ (see Figure 2e). This emission pattern corresponds to coupling out of the laser mode normal to the lattice plane with a small divergence angle of 71 and 129 mrad (see Figure S6a in the Supporting Information) along the $k_y$ and $k_x$ direction, respectively. Periodic features in the emission pattern, visible in the inset of Figure 2e, result from the finite size of the grating and are spaced by $\Delta k_{||}/k_0 = \lambda/D$, where $\lambda$ is the lasing wavelength and $D$ the array size.

To design a plasmonic-lattice laser that allows for dual-wavelength operation, we break the symmetry of the square array into a rectangular array, as recently demonstrated by Pourjamal and coworkers.[46] Depicted in eq 1, the momentum and thus the energy (eq 2) of the $[i,j] = [\pm 1,0]$ and $[i,j] = [0,\pm 1]$ bands can be independently controlled by varying $\vec{G}_1$ and $\vec{G}_2$, respectively. While the $[i,j] = [0,\pm 1]$ bands cross the $[i,j] = [\pm 1,0]$ bands at $k_x = 0$ for a square array ($|\vec{G}_x| = |\vec{G}_y|$), they do not for the rectangular array ($|\vec{G}_x| \neq |\vec{G}_y|$). Instead, the crossing of the two sets of bands with the $k_x = 0$ axis occurs at different energies (see schematic in Figure 3a). This creates two stop gaps at $k_x = 0$ where feedback is provided through second-order Bragg diffraction. With a sufficiently small difference between the lattice spacings $a_x$ and $a_y$ in $x$- and $y$-direction, we can design the rectangular lattice such that both stop gaps overlap with the QD gain spectrum. According to eq 3, lasing can then occur at two separate energies, namely $E_x \approx hc/n_{\text{eff}}a_x$ and $E_y \approx hc/n_{\text{eff}}a_y$.

We fabricated a rectangular Ag-particle lattice with $a_x = 390$ nm and $a_y = 375$ nm using template stripping and covered it with a ~180-nm-thick QD-film. Figure 3b shows the in-plane-momentum-resolved emission spectrum of the array upon excitation with a 385-nm LED. The linear $[i,j] = [1,0], [-1,0]$ bands associated to scattering by $\vec{G}_x$ form a stop gap centered at ~1.91 eV, coinciding with the stop-gap position measured for the square array ($a_x, a_y = 390$ nm). In contrast, the parabolic bands associated to diffraction by $\vec{G}_y$ form a second band opening centered at ~1.99 eV, which lies at the high-energy shoulder of the QD-emission spectrum and thus is faintly visible. The white dashed lines represent the empty lattice dispersion relation and indicate the position of the $[i,j] = [0,\pm1]$ bands.

As we pump the sample with a fs-pulsed laser at 405 nm, the QDs receive sufficient feedback at both stop gaps and we observe two emerging lasing peaks clearly separated in energy by 60 meV (see Figure 3c). The peaks at 1.93 and 1.99 eV result from second-order Bragg diffraction along the $x$- and $y$-direction, respectively. Figure 3d shows the intensity of the high- (blue line) and low-energy (red line) lasing signal as a function of pump fluence. For this sample, lasing emission occurs for both modes at nearly identical pump fluences of 0.14 mJ/cm². Although the low-energy lasing peak appears at the same energy as in the square array, the measured threshold of this sample is 0.02 mJ/cm² lower. This difference is likely related to slight differences in the quality of the QD film, which can affect the losses and therefore the lasing threshold. Compared to previous experiments, where dual-wavelength lasing was obtained from a rectangular array of lossy Ni disks[46] or a rhombohedral array of anisotropic Al particles,[47] our reported threshold values are lower by a factor ten and five, respectively. Figure 3e shows the full Fourier image of the emitted lasing signal. Compared to the square array of Figure 2, we measure a larger divergence angle of 98 and 205 mrad (see Figure S6b in the Supporting Information) along the $k_y$ and $\underline{k_x}$ direction, respectively. Possibly, the wider angle is due to small imperfections in our structure that outscatter waveguided light more strongly than the rest of the lattice and distort the outscattering pattern.

In a rectangular lattice the two lasing energies can be tuned independently. Figure 3f shows a few examples where we vary the pitch in the $y$-direction from $a_y = 375$ nm (Figure 3c) to $a_y = 385$ nm in steps of 5 nm while maintaining the spacing along the $x$-direction at $a_x = 390$ nm. The momentum-resolved emission maps highlight that the high-energy lasing peak shifts from 1.99 to 1.95 eV as $a_y$ increases, whereas the low-energy lasing emission remains at 1.93 eV.

Rectangular lattices can be used to widely tune the lasing wavelengths. The only constraint remains sufficient spectral overlap between the resonance conditions of the lattice and the gain envelope of the QDs. These restrictions could be eased at sufficiently strong pumping, because this can extend the gain window of the QDs due to high-energy emission transitions from the QD shell, as has recently been demonstrated for QD ring resonators.[29]

By filtering the lasing emission with a polarizer, we can switch between the emitted lasing wavelengths (see schematic Figure 3g). Figure 3g shows the same in-plane-momentum-resolved emission spectra as depicted in Figure 3c but polarization resolved. The measurements show that the low-energy peak associated with feedback along the $x$ direction is s-polarized (in a reference frame as shown in Figure 1c) whereas the high-energy peak corresponding to feedback along the $y$ direction is p-polarized. This polarization dependency agrees with the measurements shown in Figure 1e and confirms that feedback is provided by a TE$_0$ mode that propagates parallel to the lattice vectors with in-plane polarization. In contrast, Pourjamal and coworkers,[46] could not select between lasing wavelengths using a polarizer because the parabolic $[i,j] = [0,\pm1]$ band of their Ni arrays contained both polarization states. They ascribed this observation to quadrupolar (rather than dipolar) resonances supported by their particles. When we plot Figure 3g on a logarithmic intensity scale (see Section S4 and Figure S5 in the Supporting Information), the s-polarized emission map shows weak leakage emission from the parabolic band. However, we assign the cross-polarized signal to a finite-size effect of the lattice (see Section S4 in the Supporting Information for further discussion).

An alternative approach to achieve dual-wavelength lasing, which does not rely on a broken array symmetry, is to modify the layer stack surrounding the lattice (see Figure 4a) and receive feedback in higher-order waveguide modes. Figure 4a shows the effective waveguide mode indices for a QD film ($n_{QD} = 1.85$) surrounded by two semi-infinite epoxy layers with a high refractive index ($n_{epoxy} = 1.7$). Upon increasing the QD-layer thickness, $n_{eff}$ of the fundamental $TE_0$ and $TM_0$ modes increases from $n_{epoxy} = 1.7$ to $n_{QD} = 1.85$. If the film exceeds the cut-off thickness of ~435 nm, higher-order $TE_1$ and $TM_1$ modes are supported in addition to the fundamental modes (see inset Figure 4a for the field-intensity distributions).

To fabricate a sample that supports higher-order waveguide modes, we template-stripped a square lattice ($a_x = a_y = 355$ nm and $d = 70$ nm) with a high-index epoxy ($n_{epoxy} = 1.7$) and covered it with a ~750 nm thick film of QDs. Figure 4b shows the momentum-resolved emission spectrum of this structure upon excitation with a 385-nm LED. We observe two overlapping band structures that result from diffraction of the $TE_0$ and $TE_1$ modes by the square lattice (see white dashed and dotted lines depicting the empty-lattice dispersions of the $TE_0$ and $TE_1$ modes, respectively). As discussed above (Figure 2), the transverse-magnetic modes ($TM_0$ and $TM_1$) are hardly visible in our emission measurement because of the negligible out-of-plane polarizability of our Ag disks. As the two guided transverse-electric modes have distinct $n_{eff}$ (see Figure 4a) the band structures have energetically separated stop gaps centered at 1.91 and 2.00 eV. Using eq 3 we can deduce $n_{eff,TE_0} \approx 1.83$ and $n_{eff,TE_1} \approx 1.75$ which corresponds to a QD-film thickness of ~775 nm according to Figure 4a. Due to the tunable refractive index of the epoxy and the variable QD-film thickness, our fabrication method allows for control over the effective mode indices (Figure 4a) and therefore over the energies of the two stop bands. For this experiment, we purposely chose a higher-refractive index epoxy than in Figures 1–3, because this decreases the difference in effective index for the two TE modes and therefore decreases the energy separation between the two stop gaps. This allowed us to design the geometry such that both stop gaps overlap with the gain spectrum of the QDs.

When we excite the QD film using a fs-pulsed laser at 405 nm, we observe two lasing peaks with fwhm <2 meV (1 nm) at 1.92 and 1.97 eV in the momentum-resolved emission spectrum (see Figure 4c). Comparing the signal to the band structure in Figure 4b, it becomes apparent that the lasing emission emerges from the stop gap at $k_x = 0$ associated to the $TE_0$ mode and two points at $k_x \approx \pm 0.54\ \mu m^{-1}$ where the bands associated to $TE_0$ and $TE_1$ modes cross. At the energy where those two bands cross, a $TE_0$ mode with $\vec{k}_{TE_0}$ is diffracted into a counterpropagating $TE_1$ mode with $\vec{k}_{TE_1}$ by means of $2\vec{G}$. Coupling of the two modes gives rise to a band opening at $k_x \neq 0$, resulting in off-normal laser emission. We achieve this off-normal emission in a simple square array, without having to rely on more elaborate lattice geometries, in contrast to other recent reports.[48,49]

Figure 4d shows the output intensity of the high- (blue line) and low-energy (red line) lasing peaks as a function of pump fluence. We extract threshold values of 0.2 and 0.5 mJ/cm² for the blue and red curves, respectively. The difference in thresholds can be explained by the different overlaps of the lasing peaks with the gain window. These threshold values are higher than those we found for thin QD films (Figure 3). The difference might be related to considerable attenuation of the excitation light by the QD film, so that higher fluences are necessary for sufficient excitation of the QDs at the bottom of the film. Indeed, we estimate that the attenuation length of 405-nm light through a QD film is approximately $d = 1/\sigma\rho \approx 1\ \mu m$, where $\sigma \approx 10^{-14}\ cm^2$ is the QD absorption cross section[50] and $\rho \approx 10^{18}\ cm^{-3}$ is the QD density in the film. As the QD-film thickness approaches the attenuation length, absorption by the QD film itself is expected to negatively affect the lasing threshold. However, the threshold values are still markedly lower than for other dual-wavelength lasers recently presented.[46,47]

In summary, we have demonstrated single- and dual-wavelength lasing from plasmonic-lattice resonators with integrated QDs. Using template stripping we developed a simple method to fabricate arrays of plasmonic Ag disks on substrates with tunable refractive index. We demonstrated that the band structure and thus the resonance condition for lasing of these 2D resonators is determined by the array geometry and the layer stack surrounding the Ag particles. By drop casting a thin film of QDs onto a square array of Ag disks we obtained single-wavelength lasing. Using the full control over lattice spacings, QD-film thickness, and the substrate refractive index offered by our fabrication method, we have demonstrated two approaches for dual-wavelength lasing. First, a rectangular lattice enables tunable dual-wavelength lasing with polarization control. Second, by increasing the QD-film thickness and adjusting the refractive index of the surrounding epoxy layers we achieve off-normal lasing emission at two distinct energies from a simple square array. In this structure, in addition to lasing in fundamental waveguide modes, $TE_0$–$TE_1$ mode coupling provides feedback at a separate energy, which couples out of the lattice plane at off-normal angles. Our results highlight that films of QDs in combination with template-striped plasmonic lattices can provide a versatile platform for on-chip lasers. Based on our method, future studies could explore lattices with exotic particle shapes[47,51,52] that offer more control over the polarizability of the individual unit cell. Moreover, combining complex array geometries[41,47,48,53,54] with higher-order waveguide modes could offer an interesting route towards multiwavelength lasers at off-normal directions. Besides optimizing the array geometry, the QDs could be replaced with core/shell nanoplatelets[55] that provide higher modal gain values.[56]

ASSOCIATED CONTENT

**Supporting Information**

The Supporting Information is available free of charge on the ACS Publications website at DOI: 10.1021/acsnano.

Detailed description of our sample fabrication and optical setup (sections S1−S2); additional data supporting our analysis (sections S3−S4)

AUTHOR INFORMATION

**Corresponding Author**

*Email: dnorris@ethz.ch.

**ORCID**

Jan M. Winkler: 0000-0001-5062-8523

Max J. Ruckriegel: 0000-0002-4776-699X

Henar Rojo: 0000-0003-1543-6264

Robert C. Keitel: 0000-0002-9412-8034

Eva De Leo: 0000-0002-9677-0274

Freddy T. Rabouw: 0000-0002-4775-0859

David J. Norris: 0000-0002-3765-0678

**Present Address**

†Debye Institute for Nanomaterials Science, Utrecht University, Princetonplein 1, 3584 CC Utrecht, The Netherlands.

**Notes**

The authors declare no competing financial interest.

**Funding Sources**

This work was supported by the Swiss National Science Foundation under Award No. 200021-165559 and the European Research Council under the European Union's Seventh Framework Program (FP/2007-2013) / ERC Grant Agreement Nr. 339905 (QuaDoPS Advanced Grant). F.T.R. acknowledges support from The Netherlands Organization for Scientific Research (NWO, Rubicon Grant 680-50-1509).

ACKNOWLEDGMENTS

We thank M. Aellen, F. Antolinez, R. Brechbühler, and N. Lassaline for stimulating discussions and U. Drechsler, S. Meyer, and A. Olziersky for technical assistance.

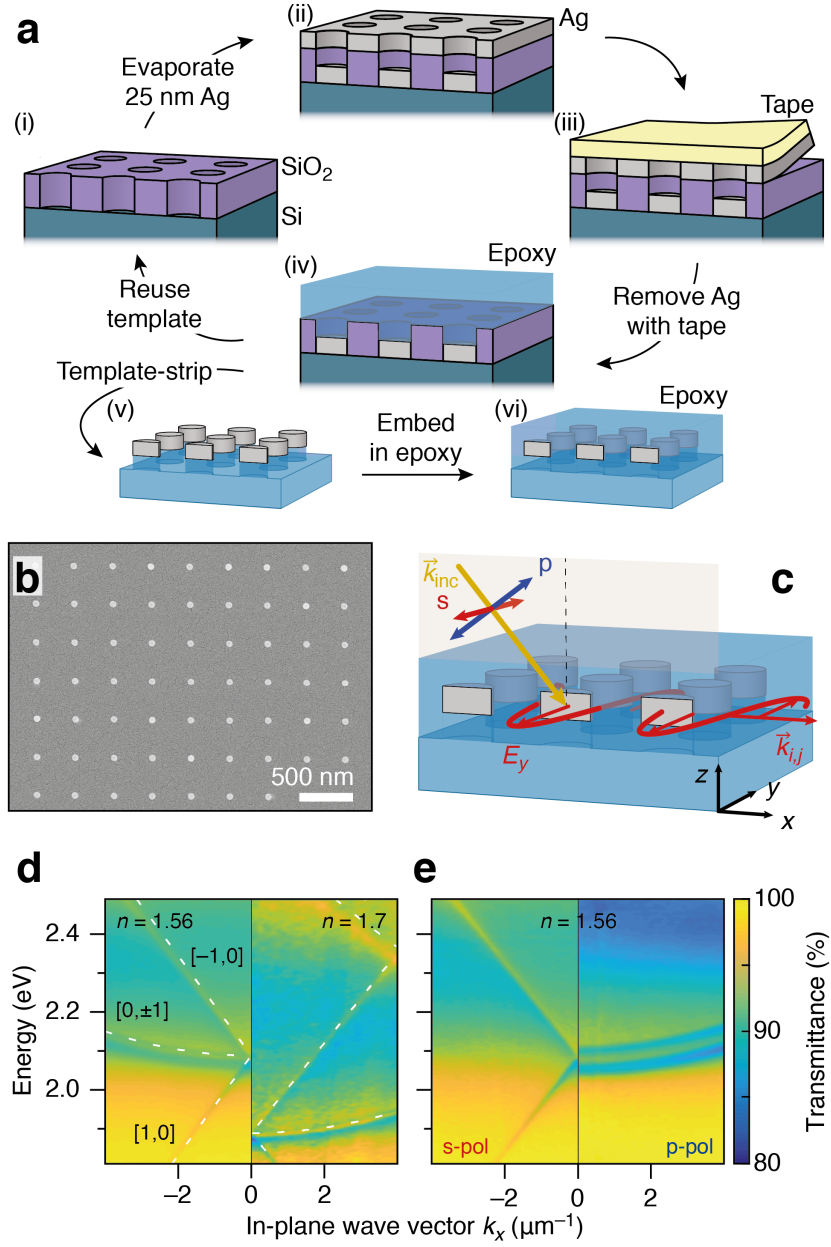

**Figure 1.** Fabrication and characterization of plasmonic lattices in a homogeneous medium. (a) Schematic overview of plasmonic-lattice fabrication via template stripping. The steps i–vi are discussed in the text. (b) Scanning electron micrograph (SEM) of a template-stripped Ag-disk lattice with pitch $a_x = a_y = 370$ nm and disk diameter $d = 60$ nm. (c) Incoming light $\vec{k}_{inc}$ with s-polarization (red arrow) is scattered into an in-plane propagating diffraction order with $\vec{k}_{i,j} = (k_x, 0, 0)$ and an electric-field along the $y$-direction. Incoming light with p-polarization cannot excite this mode because the polarizations do not match. (d) Momentum-resolved transmittance spectra of a Ag-disk lattice with $a_x = a_y = 385$ nm, $d = 70$ nm that is embedded in an epoxy with a refractive index $n = 1.56$ (left side of plot) and $n = 1.7$ (right side). The white dashed lines indicate the corresponding empty-lattice dispersions. (e) Polarization-resolved transmission map of the same plasmonic lattice as in (d), for s-polarized light (left side) and p-polarized light (right side).

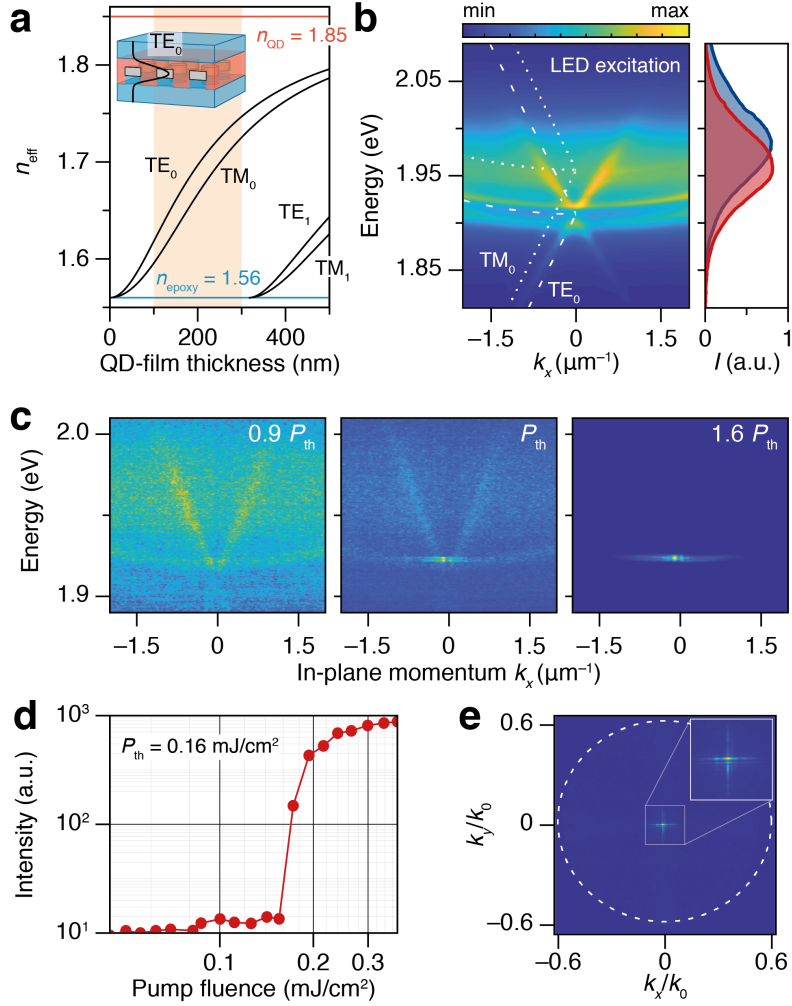

**Figure 2.** Lasing measurements of a Ag-disk square lattice covered with colloidal QDs. (a) The calculated effective mode indices $n_{eff}$ of the transverse-electric (TE) and transverse-magnetic (TM) waveguide modes supported inside the high-refractive-index QD film as a function of emitter-layer thickness. For single-wavelength-lasing experiments, film thicknesses in the orange region are targeted, below the TE$_1$/TM$_1$ cut-off thickness. Inset shows schematic of the QD film ($n_{QD} = 1.85$) integrated onto a plasmonic lattice templated stripped with epoxy of $n_{epoxy} = 1.56$. (b) Left side: below-threshold momentum-resolved emission spectrum of a ~180-nm-thick QD film placed on a plasmonic square-lattice ($a_x = a_y = 390$ nm, $d = 70$ nm). The white dashed and dotted lines depict the empty-lattice dispersion for $n_{eff,TE_0} = 1.665$ and $n_{eff,TM_0} = 1.625$, respectively. Right side: photoluminescence spectra measured from QDs in solution (blue line) and in a film (red line). (c) Momentum-resolved emission maps from a QD film on the same plasmonic lattice for increasing excitation with a fs-pulsed laser, going from below-threshold (left side; 90% of threshold fluence) to above-threshold (right side; 160% of threshold fluence). The lasing signal emerges from the upper band edge associated with the TE$_0$ mode. (d) The emission intensity from the lasing mode as a function of pump intensity, showing a threshold at 0.16 mJ/cm². (e) The Fourier image of the above-threshold emission, highlighting a low beam divergence of 71 and 129 mrad along the $k_y$ and $k_x$ direction, respectively.

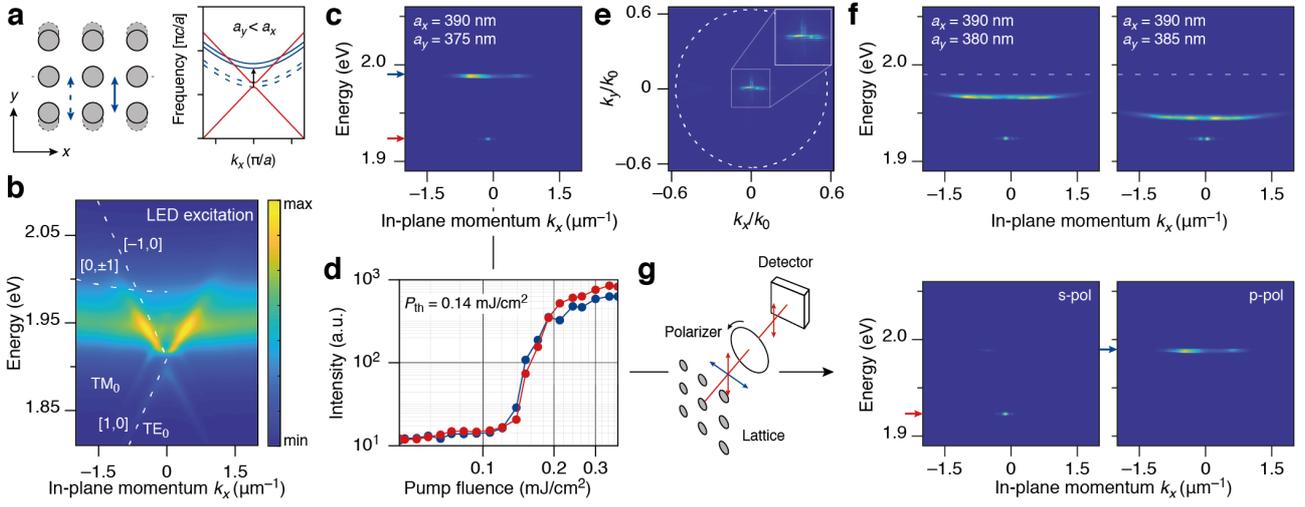

**Figure 3.** Lasing measurements of a Ag-disk rectangular lattice covered with colloidal QDs. (a) Scheme of the plasmonic lattice and its band structure when distorting from a square (dashed blue line) into a rectangular array (blue line). For a compression along the $y$-axis the parabolic $[i,j] = [0,\pm1]$ bands are shifted to higher energy. (b) Below-threshold momentum-resolved emission spectrum of a ~180-nm-thick QD film placed on a plasmonic rectangular lattice ($a_x = 390$ nm, $a_y = 375$ nm, and $d = 70$ nm). The white dashed lines depict the empty-lattice dispersion for $n_{\text{eff,TE}_0} = 1.665$. (c) Above-threshold momentum-resolved emission spectrum, showing dual-wavelength lasing at 1.99 and 1.93 eV. (d) Laser emission intensity as a function of pump fluence, indicating identical threshold values (0.14 mJ/cm²) for the high- (blue line) and low-energy signal (red line) shown in (c). (e) The Fourier image of the above-threshold emission, highlighting a low beam divergence of 98 and 205 mrad along the $k_y$ and $k_x$ direction, respectively. (f) Above-threshold momentum-resolved emission spectra for rectangular arrays with $a_x = 390$ nm, $a_y = 380$ nm and $a_x = 390$ nm, $a_y = 385$ nm, respectively. The white dashed lines indicate the position of the high-energy peak shown in (c). (g) Polarization-resolved measurements of the lasing emission of (c), for s-polarization (left side) and p-polarization (right side). We use the same linear color scale for both measurements.

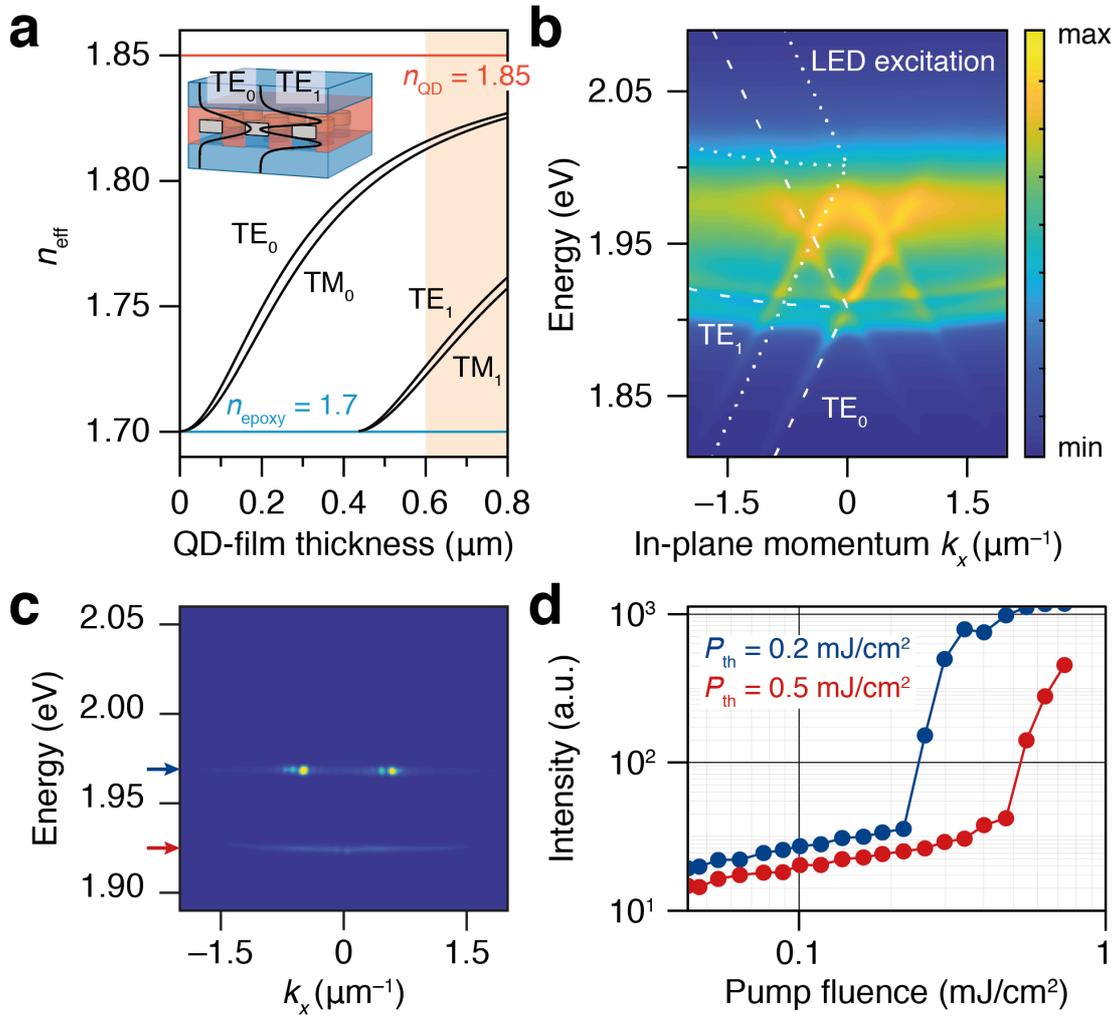

**Figure 4.** Lasing measurements of a Ag-disk square lattice covered with a thick QD film supporting higher-order waveguide modes. (a) The effective mode indices $n_{\text{eff}}$ of the transverse electric (TE) and transverse magnetic (TM) waveguide modes supported inside the high-refractive-index QD film as a function of emitter-layer thickness. In experiments for dual-wavelength lasing we target film-thicknesses above the $TE_1/TM_1$ cut-off thickness (orange region). Inset shows a schematic of the QD film ($n_{\text{QD}} = 1.85$) integrated onto a plasmonic lattice template-stripped with epoxy of $n_{\text{epoxy}} = 1.7$. (b) Below-threshold momentum-resolved emission spectrum of a ~750-nm-thick QD film placed on a plasmonic square lattice ($a_x$, $a_y = 355$ nm and $d = 70$ nm). The white dashed and dotted lines depict the empty-square-lattice dispersions for $n_{\text{eff,TE}_0} = 1.83$ and $n_{\text{eff,TE}_1} = 1.75$, respectively. (c) Above-threshold momentum-resolved emission spectrum, showing dual-wavelength lasing at 1.92 (red arrow) and 1.97 eV (blue arrow). The high-energy lasing peak is emitted at off-normal angles ($k_x \approx \pm 0.54$ μm$^{-1}$). (d) Laser emission intensity as a function of pump fluence, indicating for the high- (blue line) and low-energy signal (red line) shown in (c) threshold values of 0.2 mJ/cm$^2$ and 0.5 mJ/cm$^2$, respectively.

**Table of Contents Graphic**

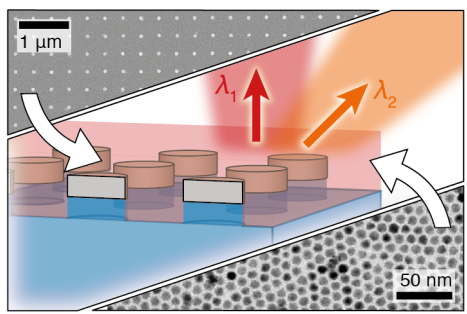